\documentclass[aps, prl, reprint, amsmath, amssymb, floatfix]{revtex4-2}

\usepackage[utf8]{inputenc}
\usepackage[T1]{fontenc}
\usepackage{microtype}          % better typography
\usepackage{graphicx}           % figures
\usepackage{xcolor}             % colours
\usepackage{amsmath, amssymb}   % math
\usepackage{bm}                 % bold math (\bm{})
\usepackage{siunitx}            % SI units  e.g. \SI{3.0e8}{\m\per\s}
\usepackage{booktabs}           % publication-quality tables
\usepackage{hyperref}           % clickable links & DOIs
\usepackage{cleveref}           % smart cross-references (\cref{})
\usepackage{natbib}

\hypersetup{
  colorlinks = true,
  linkcolor  = blue!70!black,
  citecolor  = blue!70!black,
  urlcolor   = blue!70!black,
}

\begin{document}

% ── Title & Authors ─────────────────────────────────────────
\title{%
  Splashing velocity of a viscous liquid squeezed between two parallel disks
}

\author{Navin Kumar Chandra}
\email{navin-kumar.chandra@univ-amu.fr}        % corresponding author
\affiliation{
  Aix-Marseille University, CNRS, IUSTI, Marseille 13013, France
}

\author{Pierre Perrier}
\affiliation{
  Aix-Marseille University, CNRS, IUSTI, Marseille 13013, France
}

\author{David Brutin}
\affiliation{
  Aix-Marseille University, CNRS, IUSTI, Marseille 13013, France
}

% ── Dates (optional — journals often fill these in) ─────────
% \date{\today}

% ── Abstract ────────────────────────────────────────────────
\begin{abstract}
 Squeezing of a liquid film between two approaching solid surfaces can generate a high-speed peripheral splash. Despite extensive studies on squeeze-film hydrodynamics, quantitative prediction of the splash velocity remains unresolved, with existing theory significantly overestimating experiments. Here, we present a theoretical framework to predict the ejection velocity of viscous liquid squeezed between two parallel circular disks. We show that discrepancy of theory from experiments arise due to incomplete treatment of liquid inertia and from using peak ejection velocity to represent splash velocity. By accounting for both local and convective inertia, and introducing a momentum-averaged ejection velocity, we obtain good agreement with experiments over a wide range of parameters.
\end{abstract}

\maketitle

% ============================================================

Squeezing of a liquid film between two solid surfaces is a common phenomenon encountered in everyday life, from clapping wet hands to water splashing by a vehicle's tyre on a rainy day \cite{gart2013dynamics, ge2024effect}. It is a classical problem in fluid mechanics, important for its relevance to practical applications and for its fundamental characteristics as a highly transient, spatially confined flow~\cite{moss2011highly, krassnokutski2013experimental, lang2019experimental}. Practical contexts in which squeeze-film dynamics play a central role include hydrodynamic lubrication of industrial bearings and synovial joints~\cite{kenedi1973perspectives}, the biomechanics of concussive head injury~\cite{lang2021modeling}, processing of fibre-reinforced composite materials~\cite{shuler1996transverse}, and the extensional rheology of soft matter and complex fluids~\cite{leider1974squeezing, mcclelland1983squeezing, laun1992rheometry, meeten2004squeeze, engmann2005squeeze}. The problem traces its origins to the pioneering work of Stefan~\cite{stefan1875versuche} and Reynolds~\cite{reynolds1885theory}, who analyzed the squeezing of viscous liquids in the creeping-flow regime and derived expressions for the force required to maintain a prescribed squeeze rate. Subsequent studies extended this framework to account for liquid inertia, non-Newtonian fluid properties~\cite{phan1983viscoelastic, phan1985squeeze, lang2017exact, ashkenazi2026squeeze}, and the role of slip boundary conditions at the disk surfaces~\cite{meeten2004squeeze, engmann2005squeeze}. Despite this breadth of attention, the vast majority of prior work has focused on the flow and pressure fields within the confined film. The liquid ejected radially outward from the gap, which forms a peripheral splash, has received comparatively little attention. The only systematic investigation of the ejected splash is due to Bazilevsky and Rozhkov~\cite{bazilevsky2018dome, bazilevskii2018splash, bazilevsky2020impact, bazilevskii2024round}, who combined high-speed visualization with a theoretical framework to analyze the splash formation mechanism. However, their theory significantly over-predicts the peak splash velocity, a discrepancy we quantify against our own experimental observations later in this paper. Here we identify the physical origin of this discrepancy and present a theoretical framework, validated from experiments, that yields a quantitative estimate of the splash velocity. Understanding the dynamics of such squeeze-driven liquid splashing is directly relevant to the formation of bloodstain spatter patterns encountered at crime scenes and used a source of information by forensic experts~\cite{stotesbury2016impact, faflak2021impact}.

\begin{figure*}
  \centering
  \includegraphics[width=\textwidth]{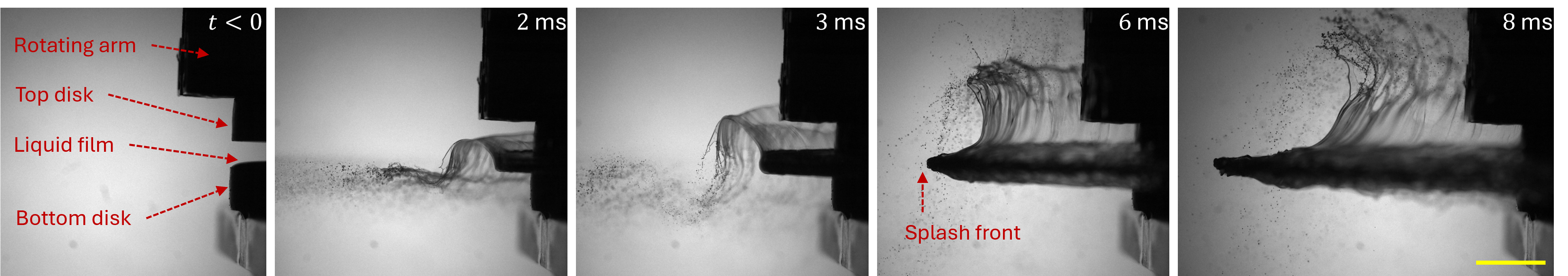}
  \caption{
  Experimental images showing the left-half of side view from high-speed visualization of splashing phenomenon. $t=0$ corresponds to the instant when the top disk touches the liquid surfaces. Scalebar shown in the last panel represents 20 mm.
  }
  \label{fig:visulization}
\end{figure*}

The term `splash,' in the present context, refers to the coherent mass of liquid observed near the periphery of the approaching disks. We generated this splash by allowing an acrylic disk of radius, $R=30$ mm to fall vertically on a thin liquid film deposited on an another disk of same radius (see Figure \ref{fig:visulization}). The top disk is rigidly mounted on a rotating lever arm, ensuring that the two disks remain parallel and axially aligned during the impact. The liquid film is created by depositing a volume of 7~mL sample which spreads in the entire top area of the bottom disk, yielding an initial film thickness $h_0 \approx 2.48$~mm. We performed experiments with three Newtonian liquids: water, and two different water-glycerol solutions referred to as G50W50 and G86W14 where 50 and 86 represents the weight percentages of glycerol in the respective solutions. The splashing phenomenon is recorded using a high-speed camera, at 5000 frames per second with a spatial resolution of $\approx0.11$ mm/pixel, positioned to capture the left-half of splash from the side view as shown in Figure~\ref{fig:visulization}. The impact velocity $u_0$ is set by releasing the top disk from different initial heights, and is estimated directly from the high-speed recordings by tracking the position of top disk's lower face prior to contact.
It is clear from Figure \ref{fig:visulization}, and also described by \citet{bazilevsky2020impact}, that the squeeze-driven splash has a complicated structure, evolving from an initial bowl shape to a dome-shaped structure before finally fragmenting into smaller droplets. It is also observed that a spray-like ejection, detached from the bulk splash, appears at the early stage of impact. This early ejecta carries only a small fraction of the total liquid mass but moves at relatively higher speed compared to the subsequent bulk splash. The exact mechanism for the formation of this early eject is not known, with one of the possibility from liquid stripped by the ambient air layer draining at high speed between the top disk and the liquid film just before contact. However, in the present work, we focus only on determining the velocity of bulk splash before its fragmentation, which can serve as a leading-order term for modelling other aspects of this process, and is more relevant to practical cases such as blood spatter at crime scenes. The splash velocity is estimated experimentally by tracking the fastest-moving splash front in the high-speed recordings, as shown in Figure \ref{fig:visulization}.

\begin{figure}[htbp]
\centering
\includegraphics[width=\columnwidth]{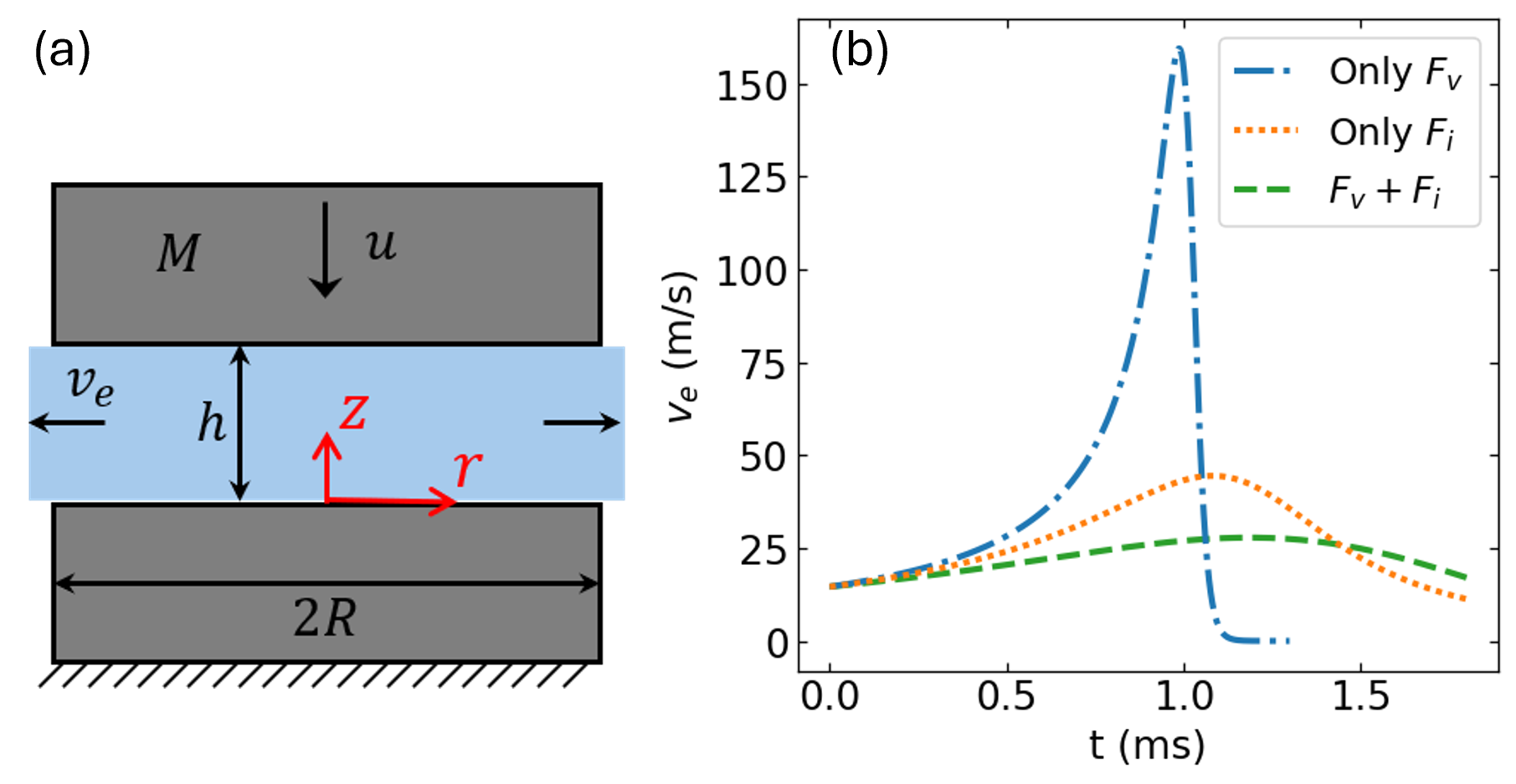}
  \caption{
    (a) Schematic representation of a viscous liquid squeezed between two parallel circular disks.
    (b) Temporal variation of depth-averaged liquid ejection velocity in the radial direction, predicted theoretically by considering values of different parameters from our experiments with G50W50 liquid, and for an initial impact velocity $u_0=2.41$ m/s.
  }
  \label{fig:schematic_theory}
\end{figure}

The key to constructing a theoretical framework for estimating the splash velocity lies in understanding the squeezing of liquid between two parallel disks, which is the origin of splash. This problem, in its simplest form, can be described with the help of a schematic diagram as shown in Figure \ref{fig:schematic_theory}a. Here, a top disk of mass $M$, and radius $R$ falls vertically with a velocity $u$ on a thin liquid film of thickness $h$, resting on top of a stationary bottom disk. The velocity field inside the liquid film can be complicated and far from being uniform \cite{lang2017exact}, however, it is easier to model the phenomenon with a depth-averaged radial velocity $v_r$ of liquid such that it depends only on $r-$coordinate, but is independent of $z-$coordinate. Assuming the liquid to be incompressible, mass conservation implies that the volume of liquid displaced by the top disk must be balanced by the flow in the radial direction, resulting in equation \ref{eq:mass_balance}.

\begin{equation}
 v_r = u \frac{r}{2h}
  \label{eq:mass_balance}
\end{equation}

The relation in equation \ref{eq:mass_balance} is a kinematic constraint and must hold true in all cases irrespective of other assumptions invoked later to model the process. The ejection velocity of liquid $v_e$, at the disk periphery that governs the splash formation, can be calculated as $v_e = v_r(r=R)$. To estimate $v_e$ one needs temporal variation of $u$ and $h$. The simplest way is to assume a constant rate of squeezing, such that $u=u_0$, and film thickness decreases linearly in time, such that $h=h_0-u_0t$. This simplification appears intuitive if the inertia of top disk is sufficiently high and remains unaffected by the squeezing process, however this approach leads to the following singularity. The assumption of constant squeezing rate introduce a timescale, $T=h_0/u_0$, which is the squeezing completion time. However, as $t\rightarrow T$, $h\rightarrow 0$ leading to the singularity $v_r\rightarrow \infty$. The resolution to this singularity comes in terms of the viscous resistance $F_v$ and the inertial resistance $F_i$ from liquid to the top disk in the axial direction which decelerate its motion. The equation of motion for the top disk is given by equation \ref{eq:disk_motion}.
\begin{equation}
 M\frac{d^2 u}{dt^2}=Mg-F_v-F_i
  \label{eq:disk_motion}
\end{equation}

Here, $g$ is the acceleration due to gravity, and this term can be neglected for most of the practical scenarios. Equation \ref{eq:disk_motion} coupled with equation \ref{eq:mass_balance}, and $u=-dh/dt$ can be solved to predict the motion of top disk and the liquid ejection velocity. \citet{bazilevsky2020impact} followed this procedure to predict the liquid ejection velocity by considering either $F_v = 3\pi\mu u R^4/2h^3$, or $F_i = \rho \pi R^4 u^2/16h^2$ only one at a time in equation \ref{eq:disk_motion}. Here, $\rho$ and $\mu$ are the density and viscosity of the liquid being squeezed. These predictions for one of our experiments with water-glycerol mixture G50W50 ($\rho \approx$ 1120 $kg/m^3$, $\mu \approx$ 5 mPa-s), initial impact velocity, $u_0 \approx$ 2.41 m/s, and mass of the top disk, $M\approx$ 518 g, are show in Figure \ref{fig:schematic_theory}b. It should be noted that $M$ in our case is not simply the mass of the acrylic disk on top, rather it is an effective mass to account for the combined moment of inertia of the disk, and the rotating lever arm on which it is mounted. The predicted peak value of liquid ejection velocity, considering only $F_v$ is $\approx$ 160 m/s, considering only $F_i$ is $\approx$ 45 m/s, which are both significantly high compared to the experimentally observed splash velocity of $\approx$ 19.5 m/s. In the rest of this manuscript we resolve this discrepancy between theory and experiments. \citet{bazilevsky2020impact} neglected $F_i$ showing that $F_i/F_v \sim 10^{-3}$, however, our calculation shows that both terms are significant and must be considered simultaneously. This is also clear from Figure \ref{fig:schematic_theory}b that the effective contribution of $F_i$ in decelerating the top disk motion is higher compared to the contribution from $F_v$, leading to smaller value of peak ejection velocity, therefore we retain both terms in our formulation. Another required correction is that, the expression for $F_i$ used in reference \cite{bazilevsky2020impact} considers only the convective acceleration of liquid, however the temporal acceleration is equally important \cite{lang2017exact}. Correct expression for $F_i$ comes from solving the momentum balance in radial direction for an axis-symmetric flow, considering only the inertial terms (including both the local and the convective inertia), as presented in the following equation \ref{eq:momentum}.

\begin{equation}
-\frac{1}{\rho}\frac{\partial p_i}{\partial r} = \frac{\partial v_r}{\partial t} + v_r \frac{\partial v_r}{\partial r}
  \label{eq:momentum}
\end{equation}

Here, $p_i$ is the liquid pressure field due to contribution from inertial terms, and it can be solved by using expression of $v_r$ from equation \ref{eq:mass_balance}. Finally, the inertial resistance in axial direction can be estimated as $F_i =\int_{0}^{R} 2\pi r p_i \,dr$, leading to the following expression in equation \ref{eq:F_i}.

\begin{equation}
F_i = \rho \pi R^4 \left( \frac{3u^2}{16h^2} + \frac{\dot{u}}{8h} \right)
  \label{eq:F_i}
\end{equation}

The dashed green curve in Figure \ref{fig:schematic_theory}b provides the prediction of liquid ejection velocity, $v_e$ for our experimental parameters described previously, and by considering both $F_v$ and corrected $F_i$ simultaneously in equation \ref{eq:disk_motion}. Here, the peak value of $v_e$ is $\approx$ 28 m/s, which is now a better prediction of the experimentally observed value ($\approx$ 19.5 m/s), compared to the other two curves presented in Figure \ref{fig:schematic_theory}b. Next, we further refine the theoretical framework to reduce the discrepancy with experiments by drawing on physical insight into the process. It can be observed from Figure \ref{fig:schematic_theory}b that the parcels of liquid ejected at different times have different velocities. This all happens during the squeezing of the thin film, as the top plates traverse a small displacement of $\approx h_0$ in the vertical direction. This should lead to a very high velocity gradient within the splash. However, internal viscous (or viscoelastic) resistance will counter this strong gradient, and it will try to homogenize the velocities among different layers of liquid in the splash. The time required for homogenization will depend on the strength of the internal resistance. This is supported by the experimental observation that a more uniform disk-like splash is formed with highly viscous and viscoelastic liquids \cite{bazilevskii2018splash, bazilevskii2024round}, while a dome-shaped splash is formed with water and other low viscosity liquids, as observed in the present experiments shown in Figure \ref{fig:visulization}a, and in Ref. \cite{bazilevsky2020impact}. Following these physical arguments, a leading order estimate for the overall splash velocity can be obtained by the conservation of linear momentum for the splash. At any instant during squeezing, the rate of radial momentum ejection is given by $\dot{\psi} = 2\pi R h \rho v_e^2$, and the momentum-averaged liquid ejection velocity can be calculated as $v_{e,avg}=\psi/m$, where $m\approx \pi R^2 h_0 \rho$ is the total mass of ejected liquid and $\psi = \int \dot{\psi}\,dt$. Using equation \ref{eq:mass_balance}, and noting that $dh = -u\,dt$, following expression (equation \ref{eq:avg_ve}) for the momentum-averaged liquid ejection velocity can be derived.

\begin{equation}
v_{e,avg}=\frac{1}{h_0} \int_{0}^{h_0} v_e \,dh
  \label{eq:avg_ve}
\end{equation}

It is worth mentioning that, given the complicated structure of the splash and the associated flow field within it, strong velocity gradients may exist not only in the axial direction but also in the radial direction. However, the momentum-averaging approach employed here is equally applicable irrespective of the direction of the velocity gradient, as long as the splash maintains a coherent structure, i.e., before its fragmentation into smaller chunks of splashes and droplets. The approach further assumes that the decrease in splash momentum during the experimental timescale, due to the external forces, namely air drag and capillary forces from any residual liquid film connecting the splash to the disks, remains small compared to the total momentum of the splash. This is justified for the present case, where splash velocities are sufficiently high. However, surface tension may become relevant at very low splash velocities, and it can be included in the force balance to estimate the transient evolution of splash velocity~\cite{bazilevskii2024round}.

\begin{figure}
\centering
\includegraphics[width= 6.9 cm]{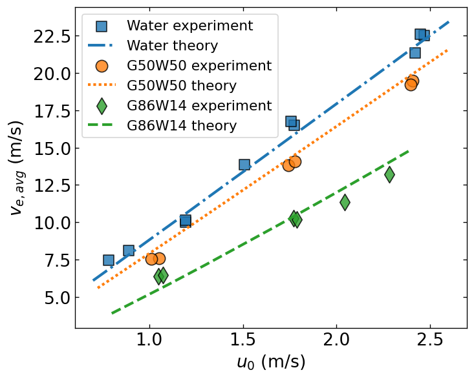}
  \caption{
  Experimental values and their corresponding theoretical predictions for the average splash velocity at different initial impact velocities for the three Newtonian liquids employed in the present work.
  }
  \label{fig:ve_vs_u0}
\end{figure}

We numerically solve equation \ref{eq:avg_ve} to compute the momentum-averaged splash velocity. For this numerical solution procedure, the initial domain considered is a liquid disk of radius $R$ and thickness $h_0$. This is subdivided into a suitable number of control volumes, each with the same radius and thickness $dh$. The value of $v_e$ for each control volume is computed from a coupled system of equations, presented in equation \ref{eq:mass_balance}, equation \ref{eq:disk_motion}, and $u = -dh/dt$, solved using the ode45 solver in MATLAB, which automatically decides the optimized number of control volumes and their respective thicknesses. The velocity of the top disk, $u$, asymptotically approaches zero, therefore, a manual stop condition on numerical solution is imposed in such a way that there is no further significant change in the total momentum of the splash. It also implies that, in theory, there is always a residual liquid layer trapped between the disks, and hence, the momentum averaging should be done considering a mass for the bulk splash that is smaller than the mass of the initial liquid film. However, as mentioned previously, we hypothesize that, due to internal resistance and cohesive forces among the liquid parcels, most of the liquid mass will be pulled into the splash, and it is justified to perform momentum averaging over the entire mass of the initial liquid film. Finally, $v_{e,avg}$ is computed using equation \ref{eq:avg_ve}. We propose that a leading order estimate for the overall splash velocity should come from $v_{e,avg}$, rather than the peak value of $v_{e}$. Figure \ref{fig:ve_vs_u0} provides a comparison of predicted splash velocity with the present experiments using water ($\mu\approx$ 1 mPa-s, $\rho\approx$ 1000 kg/$m^3$), and two different water-glycerol solutions- G50W50 ($\mu\approx$ 5 mPa-s, $\rho\approx$ 1120 kg/$m^3$) and G86W14 ($\mu\approx$ 90 mPa-s, $\rho\approx$ 1216 kg/$m^3$), for different values of initial impact velocity $u_0$. Here, error bars are not shown as they are sufficiently small to appear distinctly on the plot. The experimental values of $v_{e,avg}$ and $u_0$ are obtained by linear curve fitting to the temporal evolution of the splash front position and top disk position, respectively, and the standard error in the slope from each fit is less than $\approx$ 2.5\% at the 95\% confidence interval. Figure \ref{fig:ve_vs_u0} shows a good agreement between the proposed theory and present experiments. 

\begin{figure}[htbp]
\centering
\includegraphics[width= 6.9 cm]{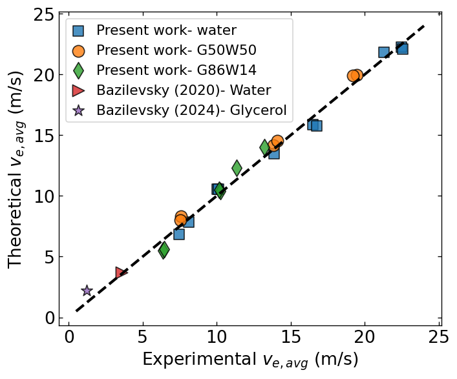}
  \caption{
  Comparison of theory and experimental values of average splash velocity, comprising all data from the present work as well as those of Bazilevsky and Rozhkov~\cite{bazilevsky2020impact, bazilevskii2024round}. The collated dataset spans nearly two orders of magnitude in disk mass $M$, one order of magnitude in disk radius $R$, and three orders of magnitude in liquid viscosity $\mu$.
  }
  \label{fig:exp_th_compare}
\end{figure}

To further ensure robustness of the proposed theoretical framework, we validate it against the splash velocities extracted from the experimental image sequences provided in reference \cite{bazilevsky2020impact} and \cite{bazilevskii2024round}. These experiments employed annular disks with a central guiding rod, instead of flat circular disks, for creating impact. Therefore, necessary geometrical modifications are done to account for the central guiding rod of radius $R_i=1$ mm, by modifying the initial domain of liquid restricted to an annular space $R_i\leq r \leq R$ in corresponding theoretical calculations. The validation of theory with these experiments from literature, as well as all the data from our own experiments are collated and presented in Figure \ref{fig:exp_th_compare}b. It is interesting to point out that these experiments taken from literature have a wide range of parameter variation compared to our experiments. For instance, mass of top disk $M$ is 3~g and 6~g compared to 518~g in the present work, disk radius $R$ is 2.5 mm compared to 30 mm in the present work, viscosity $\mu$ of liquid sample is as high as 1350 mPa-s compared to 1, 5, and 90~mPa-s in the present work. Despite this wide variation in parameter range, a good agreement with experiments (Figure \ref{fig:exp_th_compare}b) supports the robustness of the proposed theoretical framework. Apart from its robustness, it is also worth highlighting the importance of the momentum averaging approach in achieving this agreement. For the water-based experiments of Ref. \cite{bazilevsky2020impact}, the maximum splash velocity predicted from the previously employed theory, i.e., by considering only viscous or inertial resistance in isolation, is significantly high ($\sim 10^2$ m/s) compared to the actual experimental value  ($\sim 10^0$ m/s). In contrast, the prediction from the present work is close to the experimental value. This significant amount of correction underscores the importance of the momentum averaging approach in predicting the splash velocity.

In summary, we have studied the radial ejection of a viscous liquid squeezed between two parallel disks and presented a theoretical framework to predict the resulting splash velocity. We showed that the large discrepancy between previous theory and experiments stems from two distinct sources: an incomplete treatment of liquid inertia, and the assumption that the splash velocity is set by the peak ejection velocity. Accounting for both local and convective inertia, and replacing the peak ejection velocity with a momentum-averaged value as the relevant measure of splash propagation, the proposed framework yields good agreement with experiments. The proposed theory is also validated against independent literature data, confirming its robustness over a wide range of geometrical and material parameters. These results offer a physically grounded and quantitative starting point for modeling other related aspects of squeeze-driven splashing, including its morphology and subsequent fragmentation into droplets. The present framework also invites its extension to rheologically complex liquids, where viscoelastic stresses, in addition to viscous and inertial resistance, may play a significant role. Additionally, from a practical applications point of view, it would be interesting to explore the splashing from the impact of non-flat surfaces, and with flexible bottom support.

\noindent \textbf{Acknowledgment-} This work is supported by the French National Research Agency (ANR) under grant number: ANR-24-CE39-5201. The authors acknowledge support from Mr. Remy Rigard-Cerison in constructing the experimental setup.

\noindent \textbf{Data availability-} There are no publicly available research data or software supporting this manuscript. Requests for further information or data should be sent
to the authors.

% ── Bibliography ────────────────────────────────────────────
% Use BibTeX: place your .bib file in the same Overleaf project.
% revtex4-2 uses \bibliography{} — no \bibliographystyle{} needed.
\bibliography{references}

@article{reynolds1885theory,
  title={On the theory of lubrication and its application to Mr. Beauchamp Tower's experiments, including an experimental determination of the viscosity of olive oil},
  author={Reynolds, O},
  journal={Phil. Trans. Roy. Soc.},
  volume={1},
  pages={157},
  year={1885}
}

@article{stefan1875versuche,
  title={Versuche {\"u}ber die scheinbare Adh{\"a}sion},
  author={Stefan, Josef},
  journal={Annalen der Physik},
  volume={230},
  number={2},
  pages={316--318},
  year={1875},
  publisher={Wiley Online Library}
}

@article{leider1974squeezing,
  title={Squeezing flow between parallel disks. I. Theoretical analysis},
  author={Leider, Philip J and Bird, R Byron},
  journal={Industrial \& Engineering Chemistry Fundamentals},
  volume={13},
  number={4},
  pages={336--341},
  year={1974},
  publisher={ACS Publications}
}

@article{phan1983viscoelastic,
  title={Viscoelastic squeeze-film flows--Maxwell fluids},
  author={Phan-Thien, N and Tanner, RI},
  journal={Journal of Fluid Mechanics},
  volume={129},
  pages={265--281},
  year={1983},
  publisher={Cambridge University Press}
}

@article{mcclelland1983squeezing,
  title={Squeezing flow of elastic liquids},
  author={McClelland, Matthew A and Finlayson, Bruce A},
  journal={Journal of non-newtonian fluid mechanics},
  volume={13},
  number={2},
  pages={181--201},
  year={1983},
  publisher={Elsevier}
}

@article{phan1985squeeze,
  title={Squeeze film flow of ideal elastic liquids},
  author={Phan-Thien, N and Dudek, J and Boger, DV and Tirtaatmadja, V},
  journal={Journal of non-newtonian fluid mechanics},
  volume={18},
  number={3},
  pages={227--254},
  year={1985},
  publisher={Elsevier}
}

@inproceedings{laun1992rheometry,
  title={Rheometry towards complex flows: squeeze flow technique},
  author={Laun, Hans Martin},
  booktitle={Makromolekulare Chemie. Macromolecular Symposia},
  volume={56},
  number={1},
  pages={55--66},
  year={1992},
  organization={Wiley Online Library}
}

@article{shuler1996transverse,
  title={Transverse squeeze flow of concentrated aligned fibers in viscous fluids},
  author={Shuler, SF and Advani, SG},
  journal={Journal of Non-Newtonian Fluid Mechanics},
  volume={65},
  number={1},
  pages={47--74},
  year={1996},
  publisher={Elsevier}
}

@article{meeten2004squeeze,
  title={Squeeze flow of soft solids between rough surfaces},
  author={Meeten, Gerald Henry},
  journal={Rheologica acta},
  volume={43},
  number={1},
  pages={6--16},
  year={2004},
  publisher={Springer}
}

@article{engmann2005squeeze,
  title={Squeeze flow theory and applications to rheometry: A review},
  author={Engmann, Jan and Servais, Colin and Burbidge, Adam S},
  journal={Journal of non-newtonian fluid mechanics},
  volume={132},
  number={1-3},
  pages={1--27},
  year={2005},
  publisher={Elsevier}
}

@article{moss2011highly,
  title={Highly transient squeeze-film flows},
  author={Moss, EA and Krassnokutski, A and Skews, BW and Paton, RT},
  journal={Journal of fluid mechanics},
  volume={671},
  pages={384--398},
  year={2011},
  publisher={Cambridge University Press}
}

@article{krassnokutski2013experimental,
  title={An experimental study of highly transient squeeze-film flows},
  author={Krassnokutski, A and Moss, EA and Skews, BW},
  journal={Physics of Fluids},
  volume={25},
  number={6},
  year={2013},
  publisher={AIP Publishing}
}

@article{stotesbury2016impact,
  title={An Impact Velocity Device Design for Blood Spatter Pattern Generation with Considerations for High-Speed Video Analysis},
  author={Stotesbury, Theresa and Illes, Mike and Vreugdenhil, Andrew J},
  journal={Journal of forensic sciences},
  volume={61},
  number={2},
  pages={501--508},
  year={2016},
  publisher={Wiley Online Library}
}

@article{lang2017exact,
  title={Exact and approximate solutions for transient squeezing flow},
  author={Lang, Ji and Santhanam, Sridhar and Wu, Qianhong},
  journal={Physics of Fluids},
  volume={29},
  number={10},
  year={2017},
  publisher={AIP Publishing}
}

@article{bazilevsky2018dome,
  title={Dome-shaped splashes generated by the impact of a small disk on a sessile water drop},
  author={Bazilevsky, AV and Rozhkov, AN},
  journal={Physics of Fluids},
  volume={30},
  number={10},
  year={2018},
  publisher={AIP Publishing}
}

@article{bazilevskii2018splash,
  title={Splash of an elastic liquid as a rheological test of polymer solutions},
  author={Bazilevskii, AV and Rozhkov, AN},
  journal={Polymer Science, Series A},
  volume={60},
  number={3},
  pages={391--403},
  year={2018},
  publisher={Springer}
}

@article{lang2019experimental,
  title={Experimental study of transient squeezing film flow},
  author={Lang, Ji and Nathan, Rungun and Wu, Qianhong},
  journal={Journal of Fluids Engineering},
  volume={141},
  number={8},
  pages={081110},
  year={2019},
  publisher={American Society of Mechanical Engineers}
}

@article{bazilevsky2020impact,
  title={Impact of a small disk on a sessile water drop},
  author={Bazilevsky, AV and Rozhkov, AN},
  journal={Physics of Fluids},
  volume={32},
  number={8},
  year={2020},
  publisher={AIP Publishing}
}

@article{faflak2021impact,
  title={Do impact spatters depend on impact velocity, impact energy or impactor shape?},
  author={Faflak, Richard and Attinger, Daniel},
  journal={Experiments in Fluids},
  volume={62},
  number={12},
  pages={246},
  year={2021},
  publisher={Springer}
}

@article{lang2021modeling,
  title={Modeling of the transient cerebrospinal fluid flow under external impacts},
  author={Lang, Ji and Wu, Qianhong},
  journal={European Journal of Mechanics-B/Fluids},
  volume={87},
  pages={171--179},
  year={2021},
  publisher={Elsevier}
}

@article{bazilevskii2024round,
  title={Round splashes of a viscous liquid},
  author={Bazilevskii, AV and Rozhkov, AN},
  journal={Fluid Dynamics},
  volume={59},
  number={4},
  pages={756--768},
  year={2024},
  publisher={Springer}
}

@article{ashkenazi2026squeeze,
  title={Squeeze-film flow of shear-thinning fluids},
  author={Ashkenazi, Amit and Boyko, Evgeniy},
  journal={Applied Physics Letters},
  volume={128},
  number={18},
  year={2026},
  publisher={AIP Publishing}
}

@book{kenedi1973perspectives,
  title={Perspectives in Biomedical Engineering: Proceedings of a Symposium organised in association with the Biological Engineering Society and held in the University of Strathclyde, Glasgow, June 1972},
  author={Kenedi, Robert Maximilian},
  year={1973},
  publisher={Springer}
}

@article{gart2013dynamics,
  title={Dynamics of squeezing fluids: Clapping wet hands},
  author={Gart, Sean and Chang, Brian and Slama, Brice and Goodnight, Randy and Um, Soong Ho and Jung, Sunghwan},
  journal = {Phy. Rev. E},
  year={2013},
  publisher={American Physical Society}
}

@article{ge2024effect,
  title={Effect of aircraft tire wear on water spray and water displacement drag},
  author={Ge, Chenhui and Liu, Peiqing and Qu, Qiulin and Hu, Tianxiang and Zhang, Jin},
  journal={Aerospace Science and Technology},
  volume={147},
  pages={109006},
  year={2024},
  publisher={Elsevier}
}

% ── Supplemental Material (optional) ────────────────────────
% Uncomment the block below if your journal supports inline SM.
%
% \clearpage
% \setcounter{equation}{0}
% \setcounter{figure}{0}
% \setcounter{table}{0}
% \setcounter{section}{0}
% \renewcommand{\theequation}{S\arabic{equation}}
% \renewcommand{\thefigure}{S\arabic{figure}}
% \renewcommand{\thetable}{S\arabic{table}}
%
% \onecolumngrid           % SM is often single-column
% \begin{center}
%   {\large\textbf{Supplemental Material}}\\[4pt]
%   \textit{A Compelling Title That Fits on Two Lines at Most}
% \end{center}
%
% \section*{S1. Extended experimental details}
% \ldots

\end{document}